\makeatletter
\@ifundefined{@parse@version@dash}{%
	\def\@parse@version#1{\@parse@version@0#1}
	\def\@parse@version@#1/#2/#3#4#5\@nil{%
		\@parse@version@dash#1-#2-#3#4\@nil}
	\def\@parse@version@dash#1-#2-#3#4#5\@nil{%
		\if\relax#2\relax\else#1\fi#2#3#4 }
}{}
\makeatother

\documentclass[aps,pre,twocolumn,groupedaddress,longbibliography]{revtex4-2}
\usepackage{amsfonts,amssymb,amsmath} 
\usepackage{color} 
\usepackage{graphicx} 
\usepackage{epstopdf,mathrsfs} 
\usepackage[colorlinks,linktocpage,linkcolor=blue]{hyperref}

\begin{document}

\title{Langevin picture of subdiffusion in nonuniformly expanding medium}

\author{Xudong Wang$^1$}
\author{Yao Chen$^2$}
\email{ychen@njau.edu.cn}
\author{Wanli Wang$^3$}
\affiliation{$^1$School of Mathematics and Statistics, Nanjing University of Science and Technology, Nanjing, 210094, People's Republic of China \\
$^2$College of Sciences, Nanjing Agricultural University, Nanjing, 210095, People's Republic of China \\
$^3$Department of Applied Mathematics, Zhejiang University of Technology, Hangzhou 310023, People's Republic of China}

\begin{abstract}
Anomalous diffusion phenomena have been observed in many complex physical and biological systems. One significant advance recently is the physical extension of particle's motion in static medium to uniformly (and even nonuniformly) expanding medium. The dynamic mechanism of particle's motion in the nonuniformly expanding medium has only been investigated in the framework of continuous-time random walk.
To study more physical observables and supplement the theory of the expanding medium problems, we characterize the nonuniformly expanding medium with a spatial-temporal dependent scale factor $a(x,t)$, and build the Langevin picture describing the particle's motion in the nonuniformly expanding medium. By introducing a new coordinate, besides of the existing comoving and physical coordinates, we build the relation between the nonuniformly expanding medium and the uniformly expanding one, and further obtain the moments of the comoving and physical coordinates. Both exponential and power-law formed scale factor are considered to uncover the combined effects of the particle's intrinsic diffusion and the nonuniform expansion of medium. Our detailed theoretical analyses and simulations provide the foundation for studying more expanding medium problems.

\end{abstract}

\maketitle

\section{Introduction}\label{Sec1}

Beyond the classical Brownian motion, anomalous diffusion has become a very common phenomenon in the natural world, and it attracts more and more people's attention. The ordinary anomalous diffusion characterizes the nonlinear evolution of the ensemble-averaged mean-squared displacement (EAMSD)
\begin{equation}
  \langle x^2(t)\rangle \simeq 2D_\beta t^\beta
\end{equation}
with $\beta\neq1$ of the particle's motion in static medium \cite{MetzlerKlafter:2000,BarkaiGariniMetzler:2012,HoflingFranosch:2013,Manzo:2015,MogreBrownKoslover:2020}.
In recent years, however, more attention is paid to how particle diffuses in a nonstatic medium. That is to say, the medium might exhibit expanding or contracting over time, which is common in the field of biology and cosmology. The examples in biology include biological cells in interphase \cite{Alberts-etal:2015}, growing biological tissues \cite{Cowin:2004,Ambrosi-etal:2019} and lipid vesicles \cite{SzostakBartelLuisi:2001,XuHuChen:2016}. While in cosmology, the diffusion of cosmic rays in the expanding universe \cite{AloisioBerezinskyGazizov:2009} and the diffusion of fluids \cite{Haba:2014} are both worthy of investigation. Some physical processes, especially the transport process, are significantly influenced by the expanding or contracting effects of the medium. For convenience, no matter the medium is expanding or contracting, we call it the expanding medium for short in the following.

The particle's diffusion processes on the expanding medium have been partly formulated in the framework of both continuous-time random walk (CTRW) and Langevin equation. For a clear description of the change of the particle's physical coordinate $y(t)$, a new comoving coordinate $x(t)$, which is associated with a reference frame where the expanding medium appears to be static, was introduced in Ref. \cite{VotAbadYuste:2017}. The Fokker-Planck equation can be derived through the CTRW approach for unbiased normal and anomalous diffusion in uniformly expanding medium, where the expanding rate appears as the coefficient of drift term or diffusion term in macroscopic equations \cite{VotAbadYuste:2017}. The interplay between diffusive transport and the drift associated with the expanding medium gave rise to many striking effects, such as an enhanced memory of the initial condition \cite{YusteAbadEscudero:2016,VotAbadYuste:2017}, the slowing-down and even the premature halt of encounter-controlled reactions \cite{VotEscuderoAbadYuste:2018,EscuderoYusteAbadVot:2018}. These results were extended to the case including the effect of a biasing force field \cite{VotYuste:2018} or a velocity field \cite{VotAbadMetzlerYuste:2020}. Another significant advance was made by employing a generalized CTRW to deal with the general case of particles moving in domains with inhomogeneous growth and contraction rates \cite{AngstmannHenryMcGann:2017,Abad-etal:2020}.
On the other hand, the Langevin picture describing the particle's motion in uniformly expanding medium has been built, based on which, the L\'{e}vy-walk-like dynamics \cite{WangChenDeng:2019}, together with the particle's position correlation function and time-averaged mean-squared displacement (TAMSD) can be evaluated \cite{WangChen:2023}.

In fact, the CTRW and Langevin equation are the two common models to describe the anomalous diffusion. Comparing with the intuitive description of the particle's microscopic motion in CTRW, Langevin equation has the advantage of including the effect of an external force field or noises generated from a fluctuating environment \cite{CoffeyKalmykovWaldron:2004}.
Fogedby \cite{Fogedby:1994} proposed an overdamped Langevin equation in operational time $s$ coupled with a physical time process $t(s)$, named as a subordinator, so that the Langevin process's probability density function (PDF) is equal to the solution of the fractional Fokker-Planck equation of subdiffusive CTRW or L\'{e}vy flight. In the presence of an external force, the equivalence between the subordinated Langevin equations and fractional Fokker-Planck equations has also been shown in Refs. \cite{MagdziarzWeronKlafter:2008,EuleFriedrich:2009}.

Although the nonuniformly expanding medium has been discussed in the framework of CTRW \cite{AngstmannHenryMcGann:2017,Abad-etal:2020}, the advantage of Langevin equation makes it more convenient to consider the effect of an external force field, and more quantities, such as the TAMSD defined as
\cite{MetzlerJeonCherstvyBarkai:2014,BurovJeonMetzlerBarkai:2011}
\begin{equation}\label{Def-TAMSD}
  \overline{\delta^2(\Delta)}=\frac{1}{T-\Delta}\int_{0}^{T-\Delta} [x(t+\Delta)-x(t)]^2 dt.
\end{equation}
Therefore, we generalize the Langevin picture describing the particle's motion in uniformly expanding medium in Ref. \cite{WangChenDeng:2019}, and establish the one for the case with the nonuniformly expanding medium, where the scale factor is spatial-temporal dependent as $a(x,t)$.
For the convenience of theoretical analyses, we assume that the spatial variable $x$ and temporal variable $t$ are separable in the scale factor $a(x,t)$ at long-time limit. By introducing a new coordinate, besides of the comoving and physical coordinates, we can build the relation between the nonuniformly expanding medium and the uniformly expanding one, and further obtain the moments of the comoving and physical coordinates.

The remainder of this paper is organized as follows. In Sec. \ref{Sec2}, we show the detailed mathematic descriptions of the particle's dynamics in the nonuniformly expanding medium under the framework of both CTRW model and Langevin equation. Then we consider the specific subdiffusion processes in the nonuniformly expanding medium and derive the corresponding Fokker-Planck equation under the framework of Langevin equation in Secs. \ref{Sec3}. We further investigate the comoving and physical coordinates of the particles by evaluating their moments in \ref{Sec4} and \ref{Sec5}, respectively.
A summary of the main results is provided in Sec. \ref{Sec6}. In the appendices some mathematical details are collected.

\section{Nonuniformly expanding medium models}\label{Sec2}

\begin{figure}
  \centering
  \includegraphics[scale=0.75]{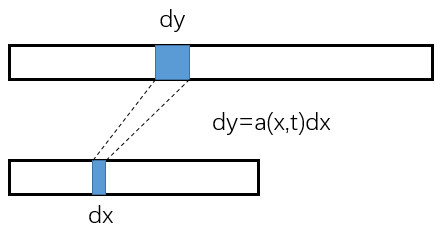}\\
  \caption{Schematic representation of the expanding medium and the change of the length of the infinitesimal cell in medium. The upper rectangle denotes the expanding medium at time $t$, while the lower one the original medium at the initial time.}\label{fig1}
\end{figure}

The scale factor is usually used to describe the extent of the expanding or contracting of the medium. For an arbitrarily changing medium, the scale factor should be spatial-temporal dependent as $a(x,t)$. Its physical meaning can be illustrated schematically in Fig. \ref{fig1}, where $dx$ is the infinitesimal width of cell $(x,x+dx)$ on the comoving coordinate. The corresponding infinitesimal cell on the physical coordinate is $(y,y+dy)$ with width $dy$. The scale factor $a(x,t)$ characterizes the degree of local expanding of medium around the comoving coordinate $x$ [i.e., the infinitesimal cell $(x,x+dx)$] at time $t$.
The spatial-temporal dependent scale factor $a(x,t)$ reduces to the temporal dependent one $a(t)$ when the medium expands or contracts at a homogeneous rate. At any fixed time $t$, $a(x,t)>1$ implies the expanding scenario around coordinate $x$, while $a(x,t)<1$ means the contracting case. The nonuniformly expanding medium can present both expanding and contracting scenarios in different places at the same time due to the spatial dependence.

In CTRW models, the particle's motion consists of a series of waiting times and jump lengths, where the particle stays in some certain position for a random time $\Delta t$ at the waiting state and performs the instantaneous jump with random length $\Delta y$ at the jump state \cite{MetzlerKlafter:2000,KlafterSokolov:2011}. The diffusing particles stick to the expanding medium and experience a drift even when they stay in the phase of waiting time. While at the moment of jumping event, the actual particle's physical position is also influenced by the expanding of medium.

The difficulty of characterizing the particle's motion on the expanding medium lies in the fact that the particle's physical position changes even at the waiting states as the medium expands. To solve this problem, a new comoving coordinate $x(t)$, i.e., spatial coordinates
referring to the initial fixed domain of medium, is introduced in Refs. \cite{VotAbadYuste:2017,AngstmannHenryMcGann:2017,VotYuste:2018,Abad-etal:2020,VotAbadMetzlerYuste:2020,WangChen:2023}. If the medium expands uniformly with the scale factor $a(t)>0$, then the physical coordinate $y(t)$ and comoving coordinate $x(t)$ are related through an equality for any physical time $t$: \cite{VotAbadYuste:2017,VotYuste:2018,VotAbadMetzlerYuste:2020,WangChen:2023}
\begin{equation}\label{Eq-at1}
  y(t)=a(t)x(t).
\end{equation}
The initial condition of the scale factor is $a(0)=1$, which implies the medium has not yet expanded and the two coordinates coincide at time $t=0$.
The advantage of introducing comoving coordinate $x(t)$ is that the descriptions of both waiting states and jumping states are effective with respect to the comoving coordinate. More precisely, at waiting states, the particle keeps still with respect to the expanding medium itself. In other words, its comoving coordinate $x(t)$ does not change. While at jumping states, letting the jump length PDF $w_y(\Delta y)$ describe the intrinsic random motion of the particle, the corresponding jump length with respect to the comoving coordinate is
\begin{equation}\label{Eq-at2}
  \Delta x=\Delta y/a(t)
\end{equation}
for uniformly expanding medium.

More generally, the medium might expands nonuniformly, and Eqs. \eqref{Eq-at1} and \eqref{Eq-at2} do not hold any more, which have been investigated in the framework of CTRW \cite{AngstmannHenryMcGann:2017,Abad-etal:2020}.
Our aim is to extend the procedures in CTRW framework to Langevin equation, which describes the particle's motion by a stochastic differential equation of physical coordinate $y(t)$. Now, the scale factor is not temporal dependent as $a(t)$, but a spatial-temporal dependent one, denoted by $a(x,t)$, the initial condition of which is $a(x,0)=1$ for any $x$. As Fig. \ref{fig1} shows, it holds that $dy = a(x,t) dx$. Then we perform the integral on both sides, and obtain
\begin{equation}\label{Eq-axt1}
   y=\int_0^x a(x',t)dx'=:g(x,t).
\end{equation}
Compared with Eq. \eqref{Eq-at1}, the physical coordinate $y$ is not the product of the scale factor and the comoving coordinate, but the integral of the $a(x,t)$ over comoving coordinate $x$. Since $y$ depends on both $x$ and time $t$, we denote it as a new function $g(x,t)$ in Eq. \eqref{Eq-axt1}, so that the comoving coordinate can be expressed as $x=g^{-1}(y,t)$.

We have built the mapping between two coordinate systems through Eq. \eqref{Eq-axt1}. The next step is to characterize the particle's motion through the comoving coordinate. For a clear physical interpretation, we start from the CTRW model with waiting state and jump state. In the waiting state, the particle's comoving coordinate $x$ does not change. While in the jump state, the particle makes a jump from coordinate $y$ to $y+\Delta y$ with length $\Delta y$ at time $t$. Then the corresponding change in comoving coordinate is from $x$ to $x+\Delta x$. Considering the relation in Eq. \eqref{Eq-axt1}, we have
\begin{equation}\label{Eq-axt2}
  \Delta x= g^{-1}(y+\Delta y,t)-g^{-1}(y,t).
\end{equation}
Performing a Taylor expansion on the right-hand side around the point $y$, we arrive at
\begin{equation}\label{Eq-axt3}
  \Delta x= \Delta y /a(x,t)+o(\Delta y),
\end{equation}
where the result
\begin{equation}
  \frac{\partial g^{-1}(y,t)}{\partial y}=\left(\frac{\partial g(x,t)}{\partial x}\right)^{-1}=\frac{1}{a(x,t)}
\end{equation}
has been used. The notation $o(\Delta y)$ denotes the quantity of the higher order than $\Delta y$.

In the form of Langevin equation, the particle's trajectory can be approximated by the cumulative sum of increments $\Delta y$ in each time interval $[t,t+\Delta t]$. In this approximation, the particle's motion in time interval $[t,t+\Delta t]$ can be understood as the combination of a waiting time $\Delta t$ and a jump length $\Delta y$. Based on this understanding, by dividing $\Delta t$ on both sides of Eq. \eqref{Eq-axt3} and letting $\Delta t\rightarrow0$, we obtain $\lim_{\Delta t\rightarrow0}\Delta y/\Delta t=dy_I(t)/dt$ and $\lim_{\Delta t\rightarrow0}o(\Delta y)/\Delta t=0$ on the right-hand side. Here the intrinsic displacement $y_I(t_n)=\sum_{i=1}^n \Delta y_i$ denotes the particle's position at time $t=t_n$ in a static medium, which is different from the physical coordinate $y(t)$ in Eq. \eqref{Eq-axt1} for the case with a nonuniformly expanding medium. In other words, Eq. \eqref{Eq-axt3} implies that the Langevin equations of comoving coordinate $x(t)$ and intrinsic displacement $y_I(t)$ only differs by the scale factor $a(x,t)$, i.e.,
\begin{equation}\label{Eq-axt2}
  \frac{d}{dt}x(t)=\frac{1}{a(x,t)}\frac{d}{dt}y_I(t).
\end{equation}

Equations \eqref{Eq-axt1} and \eqref{Eq-axt2} are the main results of building the Langevin picture of particle's motion on the nonuniformly expanding medium with the scale factor $a(x,t)$. For the particle undergoing a specific motion, such as Brownian motion, subdiffusive CTRW and other anomalous diffusion processes, the $y_I(t)$ in the Langevin equation can be given in corresponding forms. Based on Eq. \eqref{Eq-axt2}, the Langevin equation of the comoving coordinate $x(t)$ can be obtained, and the Fokker-Planck equation and the moments of the comoving coordinate $x(t)$ can be derived. Further, the results over the comoving coordinate $x(t)$ can be turned into the physical coordinate $y(t)$ through Eq. \eqref{Eq-axt1}.
Actually, the description in Eq. \eqref{Eq-axt1} is similar to the one built in the CTRW framework \cite{AngstmannHenryMcGann:2017,Abad-etal:2020} [the scale factor is denoted by $\nu(x,t)$ there]. The equivalence between the models built by Langevin equation here and those by CTRW in Refs. \cite{AngstmannHenryMcGann:2017,Abad-etal:2020} can be verified through the Fokker-Planck equation governing the PDF of $x(t)$ in the next section.

\section{Subdiffusion in nonuniform expanding medium}\label{Sec3}

Let us consider the subdiffusion process described by a subordinated overdamped Langevin equation: \cite{Fogedby:1994,BauleFriedrich:2005,BauleFriedrich:2007,Magdziarz:2009,ChechkinSokolov:2021}
\begin{equation}\label{model1}
\frac{d}{ds}{y}_I(s)=\sqrt{2D}\xi(s), \quad \frac{d}{ds}{t}(s)=\eta(s),
\end{equation}
where $D$ is the constant diffusivity, $\xi(s)$ is the Gaussian white noise with zero mean value and correlation function $\langle \xi(s_1)\xi(s_2) \rangle=\delta(s_1-s_2)$. The L\'{e}vy noise $\eta(s)$, regarded as the formal derivative of the $\alpha$-stable subordinator $t(s)$, is independent of the Gaussian white noise $\xi(s)$. The characteristic function of the subordinator $t(s)$ with $0<\alpha<1$ is \cite{BauleFriedrich:2005,WangChenDeng:2019,ChenWangDeng:2019-3}
\begin{equation}\label{char}
  \langle e^{-\lambda t(s)}\rangle=e^{-s \lambda^\alpha}.
\end{equation}
There are two time variables in the coupled Langevin equation, physical time $t$ and operational time $s$. The notation $y_I(s)$ denotes the particle's intrinsic displacement over operational time $s$ without an expanding medium. The corresponding displacement over physical time $t$ is denoted as a compound process $y_I(t):=y_I(s(t))$, where $s(t)$ is the corresponding inverse subordinator, defined by
\begin{equation}
  s(t)=\inf_{s>0}\{s:t(s)>t\}.
\end{equation}

Equation \eqref{model1} describes the particle's intrinsic motion without considering an expanding medium. The key of considering the effect of the nonuniformly expanding medium is to build the Langevin equation of comoving coordinate $x(t)$ by using the relation in Eq. \eqref{Eq-axt2}. Since Eq. \eqref{Eq-axt2} is valid on physical time $t$, we merge the two sub-equations in Eq. \eqref{model1} into one equation over physical time $t$, which is
\begin{equation}\label{modelyt}
  \frac{d}{dt}{y}_I(t)=\sqrt{2D}\bar\xi(t),
\end{equation}
where the noise $\bar\xi(t)$ is defined as \cite{CairoliBaule:2015-2,ChenWangDeng:2019-2}
\begin{equation}\label{Def-newnoise}
  \bar\xi(t):=\frac{dB[s(t)]}{dt}=\xi[s(t)]\frac{ds(t)}{dt},
\end{equation}
and $B(\cdot)$ is the standard Brownian motion.
Equation \eqref{modelyt} is obtained by replacing $s$ by the inverse subordinator $s(t)$ in the first equation of Eq. \eqref{model1}.
Therefore, for the subdiffusion process in the nonuniformly expanding medium with the scale factor $a(x,t)$, the Langevin equation with respect to the comoving coordinate $x(t)$ is
\begin{equation}\label{modelxt}
\frac{d}{dt}{x}(t)=\frac{\sqrt{2D}}{a(x,t)}\bar\xi(t).
\end{equation}

It should be noted that due to the spatial-dependent $a(x,t)$ in Eq. \eqref{modelxt}, this Langevin equation may take different interpretations, It\^{o}, Stratonovich, or H\"{a}nggi-Klimontovich over the multiplicative noise \cite{Gardiner:1983,Ito:1944,Stratonovich:1966,Hanggi:1982,Klimontovich:1990,KuboTodaHashitsume:1985}. Generally, the interpretation of integration with respect to the multiplicative noise is related to the examined process and the nature of the noise \cite{Gardiner:1983,KuboTodaHashitsume:1985,WongZakai:1965,VolpeWehr:2016}. Different interpretation leads to different processes $x(t)$ and the corresponding Fokker-Planck equations.
Let $W(x,t)$ be the PDF of finding the particle's comoving coordinate $x$ at time $t$. When $\alpha=1$, i.e., $\bar{\xi}(t)$ recovers the Gaussian white noise $\xi(t)$, the Fokker-Planck equation governing $W(x,t)$ reads \cite{LeibovichBarkai:2019,CherstvyChechkinMetzler:2013}
\begin{equation}\label{FKE1}
 \frac{\partial W(x,t)}{\partial t}=2D\frac{\partial}{\partial x} \left\{a(x,t)^{A-2}\frac{\partial}{\partial x}\Big[a(x,t)^{-A} W(x,t)\Big]\right\},
\end{equation}
where $A = 0$ for H\"{a}nggi-Klimontovich, $A = 1$ for Stratonovich, or $A = 2$ for It\^{o} interpretation. Corresponding to each value of $A$ in the Fokker-Planck operator in Eq. \eqref{FKE1}, the different interpretation of the integration can be transformed into the It\^{o} sense with different drift term in Langevin equation \eqref{modelxt} \cite{LeibovichBarkai:2019}. Therefore, the results derived in the It\^{o} sense can be extended to the Stratonovich sense or H\"{a}nggi-Klimontovich sense by replacing the Fokker-Planck operator, such as deriving Fokker-Planck equation and Feynman-Kac equation \cite{WangChenDeng:2018,CairoliBaule:2017}.
In the following, we interpret the Langevin equation \eqref{modelxt} in the Stratonovich sense to ensure the correct limiting transition for the noise with infinitely short correlation times \cite{WestBulsaraLindenbergSeshadriShuler:1979,CherstvyChechkinMetzler:2013}. That is to say, $A=1$ and the Fokker-Planck equation with $\alpha=1$ is
\begin{equation}\label{FKE2}
 \frac{\partial W(x,t)}{\partial t}=2D\frac{\partial}{\partial x} \left\{\frac{1}{a(x,t)}\frac{\partial}{\partial x}\left[\frac{1}{a(x,t)} W(x,t)\right]\right\}.
\end{equation}
When $0<\alpha<1$, the corresponding Fokker-Planck equation will contain the Niemann-Liouville fractional derivative: \cite{SokolovKlafter:2006,SokolovKlafter:2006,MagdziarzWeronKlafter:2008,Magdziarz:2009,EuleFriedrich:2009,CairoliBaule:2017,WangChenDeng:2018,ChenWangDeng:2019-2}
\begin{equation}\label{FKE3}
\begin{split}
\frac{\partial W(x,t)}{\partial t}=2D\frac{\partial}{\partial x} \left\{\frac{1}{a(x,t)}\frac{\partial}{\partial x}\left[\frac{1}{a(x,t)} D_t^{1-\alpha}W(x,t)\right]\right\}.
\end{split}
\end{equation}
The most direct way of obtaining Eq. \eqref{FKE3} is to take the Laplace symbol $p=0$ in Feynman-Kac equations of Ref. \cite{CairoliBaule:2017,WangChenDeng:2018} and to turn the Fokker-Planck operator to the one in Stratonovich sense. The consistence between Eqs. \eqref{FKE2}, \eqref{FKE3} and Eqs. (43), (82) of Ref. \cite{Abad-etal:2020} [by removing the reaction term there] derived under the CTRW framework implies that our Langevin picture of subdiffusion process in the nonuniformly expanding medium is effective.

Next, based on the Langevin picture built in Eqs. \eqref{Eq-axt1} and \eqref{Eq-axt2}, we can investigate the long-time limit of the moments of the particle's displacement over both the comoving and physical coordinates. The leading terms of the moments at short-time limit will be consistent to those in the case of a static medium. More detailed analyses about the short-time limit, such as the sub-leading terms and the strong dependence on the initial condition, can be found in Ref. \cite{Abad-etal:2020}. The long-time behavior of the scale factor $a(x,t)$ plays the determining role when considering the long-time limit of the moments of the diffusion process on the expanding medium. Therefore,
for convenience, we assume that the scale factor is separable with respect to its spatial and temporal variables at long-time limit, i.e.,
\begin{equation}\label{separable}
  a(x,t)\simeq a_1(x)a_2(t).
\end{equation}
Then at long-time limit, the Langevin equation \eqref{modelxt} of the comoving coordinate $x(t)$ behaves as
\begin{equation}\label{modelxt2}
\frac{d}{dt}{x}(t)=\frac{\sqrt{2D}}{a_1(x)a_2(t)}\bar\xi(t).
\end{equation}
Turning the factor $a_1(x)$ to the left and introducing a new variable
\begin{equation}\label{z-x}
  z=\int_0^x a_1(x')dx',
\end{equation}
we obtain the Langevin equation of $z(t)$:
\begin{equation}\label{modelzt}
  \frac{d}{dt}z(t)=\frac{\sqrt{2D}}{a_2(t)}\bar\xi(t).
\end{equation}
On the other hand, based on Eq. \eqref{Eq-axt1}, we can build the relation between the physical coordinate $y(t)$ and the new variable $z(t)$:
\begin{equation}\label{modelyzt}
  y(t)=\int_0^x a_1(x')a_2(t)dx'=a_2(t)z(t).
\end{equation}

It is interesting to find that the Eqs. \eqref{modelzt} and \eqref{modelyzt} are consistent to the Langevin picture of uniformly expanding medium established in Ref. \cite{WangChen:2023}, where $z(t)$ and $y(t)$ are the corresponding comoving and physical coordinates with respect to the uniformly expanding medium with the scale factor $a_2(t)$. In other words, only if the scale factor $a(x,t)$ of the nonuniformly expanding medium is separable at long-time limit as Eq. \eqref{separable} shows, the nonuniformly expanding medium problems can be turned into the uniform version.
Therefore, we can directly obtain the second moments of $z(t)$ and $y(t)$ by using the results in Ref. \cite{WangChen:2023}.

By contrast, the moments of the comoving coordinate $x(t)$ cannot be obtained easily due to the spatial dependent scale factor $a(x,t)$. In general, the EAMSD of $x(t)$ can be derived through two ways. The first one is based on the Fokker-Planck equation governing the PDF $W(x,t)$ in Eq. \eqref{FKE3}, and the another one is turning the moments of $z(t)$ to those of $x(t)$ by utilizing the nonlinear relation in Eq. \eqref{z-x}.
Next, we will take specific scale factor $a(x,t)$ to evaluate the moments of $x(t)$ and $y(t)$, respectively.

\section{Comoving coordinate of particle in expanding medium}\label{Sec4}

Assume that the scale factor $a(x,t)$ is separable at long-time limit, i.e., Eq. \eqref{separable} is valid. The temporal dependent part $a_2(t)$ will be discussed in two different forms, exponential and power-law forms, respectively.
We first investigate the exponential form $a_2(t)=e^{\gamma t}~(\gamma>0)$. Multiplying both sides of Eq. \eqref{FKE3} by $x^2$ and integrating over $x$, we obtain
\begin{equation}
  \frac{\partial}{\partial t}\langle x^2(t)\rangle=
  4De^{-2\gamma t}  D_t^{1-\alpha}\langle I[x(t)]\rangle,
\end{equation}
where the integration by parts has been used twice and the boundary terms vanish, and
\begin{equation}
  I(x)=\frac{a_1(x)-xa_1'(x)}{a_1^3(x)}.
\end{equation}
Performing the Laplace transform $(t\rightarrow\lambda)$ yields
\begin{equation}
  \lambda \langle x^2(\lambda)\rangle = 4D (\lambda+2\gamma)^{1-\alpha}\langle I[x(\lambda+2\gamma)]\rangle,
\end{equation}
where we assume that the particles start the motion from the origin so that the initial term in Laplace transform vanishes. Then we consider the asymptotics with small $\lambda$ (i.e., large $t$), and obtain
\begin{equation}\label{EAMSD1}
  \lambda \langle x^2(\lambda)\rangle \simeq 4D(2\gamma)^{1-\alpha} \langle I[x(2\gamma)]\rangle.
\end{equation}
The right-hand side is a constant independent of $\lambda$, though the exact value of $\langle I[x(2\gamma)]\rangle$ is unknown for general $a_1(x)$. Performing the inverse Laplace transform on Eq. \eqref{EAMSD1}, we know that the EAMSD of the comoving coordinate $x(t)$ tends to a constant for any form of $a_1(x)$, i.e.,
\begin{equation}\label{EAMSDx1}
  \langle x^2(t)\rangle \simeq C,
\end{equation}
presenting the similar result to the case with uniformly expanding medium \cite{WangChen:2023}.

If the $a_2(t)$ is in the power-law form $a_2(t)=t^\gamma~(\gamma>0)$, then the Laplace transform does not work any more. Instead, we calculate the moments of $z(t)$ through the time rescaling approach first, and then turn the moments to the ones of $x(t)$. In detail, the Langevin equation \eqref{modelzt} of $z(t)$ becomes
\begin{equation}\label{modelzt2}
  \frac{d}{dt}z(t)=\sqrt{2D} t^{-\gamma} \bar\xi(t)
\end{equation}
for power-law formed $a_2(t)$. When $\alpha=1$, the compound noise $\bar\xi(t)$ recovers to the Gaussian white noise $\xi(t)$, and the intrinsic subdiffusion process returns back to Brownian motion. Then $z(t)$ in Eq. \eqref{modelzt2} is the form of scaled Brownian motion, i.e., $z(t)\overset{d}{=}B(t^{1-2\gamma})$, which has been investigated in many references \cite{LimMuniandy:2002,ThielSokolov:2014,JeonChechkinMetzler:2014,Safdari-etal:2015,SposiniMetzlerOshanin:2019}. The notation $\overset{d}{=}$ denotes the identical distribution.
Now for the subdiffusion process with $\alpha<1$, we employ the time rescaling approach and get
\begin{equation}\label{ZSBM}
  z(t)\overset{d}{=}\sqrt{2D}\mu^{-\frac{1}{2}}B[s(t^\mu)],
\end{equation}
where the derivations are presented in Appendix \ref{App1}. Note that Eq. \eqref{ZSBM} holds in the condition that $\mu=1-2\gamma/\alpha>0$, i.e.,
\begin{equation}
  \gamma<\alpha/2.
\end{equation}
Then we utilize the self-similarity of the inverse subordinator and Brownian motion that \cite{BeckerMeerschaertScheffler:2004}
\begin{equation}
  s(t)\overset{d}{=}t^\alpha s(1), \quad B(t)\overset{d}{=}t^{1/2} B(1),
\end{equation}
and obtain
\begin{equation}\label{ztd}
  z(t)\overset{d}{=}\sqrt{2D}\mu^{-\frac{1}{2}}B[s(1)]t^{\mu\alpha/2}.
\end{equation}
Based on Eq. \eqref{ztd}, we can get the moments of $z(t)$ of arbitrary order.

The relation between $z(t)$ and $x(t)$ is built in Eq. \eqref{z-x}, which depends on the specific form of $a_1(x)$. For convenience, we consider the power-law form $a_1(x)=|x|^\beta~(\beta>-1)$. Then $z=|x|^{1+\beta}\mathrm{sgn}(x)/(1+\beta)$, and
\begin{equation}\label{EAMSDx3}
  \langle x^2(t)\rangle \propto t^{\frac{\alpha-2\gamma}{1+\beta}}.
\end{equation}
The detailed derivation and the coefficient of Eq. \eqref{EAMSDx3} are presented in Appendix \ref{App2}. Note that the condition $\beta>-1$ is to ensure the growth condition for existence and uniqueness of the solution of the Langevin equation \eqref{modelxt2} with multiplicative noise \cite{Oksendal:2005,Gardiner:1983}.

If $\gamma>\alpha/2$, then the time rescaling approach in Eq. \eqref{ZSBM} is not effective. Actually, the coordinate $z(t)$ behaves as the comoving coordinate with respect to the uniformly expanding medium with scale factor $t^{-\gamma}$, and its EAMSD has been obtained in Ref. \cite{WangChen:2023}:
\begin{equation}
  \langle z^2(t)\rangle \simeq \frac{2D\Gamma(2\gamma-\alpha)}{\Gamma(2\gamma)},
\end{equation}
presenting a localization diffusion. The time-dependent diffusivity $t^{-\gamma}$ in Eq. \eqref{modelzt2} decays rapidly so that the EAMSD does not increase at long-time limit, similar to the particles in a harmonic potential \cite{Jeon.etal:2011,JeonMetzler:2012,JeonChechkinMetzler:2014,WangChenDeng:2020-2,WangChen:2023}. Therefore, the distribution of $z(t)$ tends to be temporal-independent at long-time limit, so does the comoving coordinate $x(t)$ due to the relation in Eq. \eqref{z-x}. That is to say, the particles exhibit the localization diffusion with respect to the comoving coordinate $x(t)$ as Eq. \eqref{EAMSDx1} shows for any form of $a_1(x)$ in the case of $\gamma>\alpha/2$.

\section{Physical coordinate of particle in expanding medium}\label{Sec5}

The moments of particle's EAMSD with respect to the physical coordinate $y(t)$ can be obtained theoretically based on the moments of the comoving coordinate $x(t)$ and the relation between them in Eq. \eqref{Eq-axt1}. Fortunately, there is a direct way to get around the difficulty of evaluating the moments of the comoving coordinate $x(t)$ in Sec. \ref{Sec4}.
As the last two paragraphs in Sec. \ref{Sec3} says, the Eqs. \eqref{modelzt} and \eqref{modelyzt} are consistent to the Langevin picture of the uniformly expanding medium established in Ref. \cite{WangChen:2023}, where $z(t)$ and $y(t)$ are the corresponding comoving and physical coordinates with respect to the uniformly expanding medium with the scale factor $a_2(t)$. We only present the moments of the physical coordinate $y(t)$ here. The corresponding Fokker-Planck equation governing the PDF of $y(t)$ can be derived based on the relation in Eq. \eqref{Eq-axt1} and chain rules. The explicit form contains a comoving fractional derivative and looks more complex than the one of $x(t)$, which has been obtained in the CTRW framework in Ref. \cite{Abad-etal:2020}.

The noise $\bar\xi(t)$ in Eq. \eqref{modelzt} characterizes the intrinsic subdiffusive motion. More generally, the noise $\bar\xi(t)$ can be extended to the L\'{e}vy noise and the velocity process, to characterize the intrinsic L\'{e}vy flight \cite{ShlesingerZaslavskyFrisch:1995,BouchaudGeorges:1990,VahabiSchulzShokriMetzler:2013,FroembergBarkai:2013-2} and L\'{e}vy-walk-like dynamics \cite{KlafterBlumenShlesinger:1987,ZumofenKlafter:1993,ZaburdaevDenisovKlafter:2015,WangChenDeng:2019,ChenWangDeng:2019-3} in the nonuniformly expanding medium.
Here we only show the results corresponding the intrinsic subdiffusive motion characterized by our model in Eq. \eqref{model1}.
For the consistence to the results in the case of the uniformly expanding medium, we select specific scale factor $a(x,t)$, so that its  temporal dependent part $a_2(t)$ has the same form as the one in Ref. \cite{WangChen:2023}, i.e.,
\begin{equation}\label{Sim-a2t1}
  a_2(t)=e^{\gamma t}, \quad  \gamma>0,
\end{equation}
for exponentially expanding medium, and
\begin{equation}\label{Sim-a2t2}
  a_2(t)=(1+t/t_0)^\gamma, \quad  \gamma>0,
\end{equation}
for power-law-formed expanding medium. For the exponential-formed $a_2(t)$, the theoretical EAMSD of the physical coordinate $y(t)$ is
\begin{equation}\label{EAMSDy1}
  \langle y^2(t)\rangle \simeq 2^{1-\alpha}\gamma^{-\alpha}De^{2\gamma t},
\end{equation}
and the ensemble-averaged TAMSD of the physical coordinate $y(t)$ is
\begin{equation}\label{TAMSDy1}
  \langle\overline{\delta^2(\Delta)}\rangle \simeq \frac{4D}{(2\gamma)^{1+\alpha}}T^{-1}e^{2\gamma T}.
\end{equation}
While for the power-law-formed $a_2(t)$, it holds that
\begin{equation}\label{EAMSDy2}
  \begin{split}
    \langle y^2(t)\rangle\simeq\left\{
    \begin{array}{ll}
          \frac{2Dt_0^{\alpha-2\gamma}\Gamma(2\gamma-\alpha)}{\Gamma(2\gamma)}t^{2\gamma}, &~ \gamma>\frac{\alpha}{2}, \\[5pt]
      \frac{2D}{(\alpha-2\gamma)\Gamma(\alpha)}t^{\alpha}, & ~\gamma<\frac{\alpha}{2},
    \end{array}\right.
  \end{split}
\end{equation}
and
\begin{equation}\label{TAMSDy2}
\begin{split}
    \langle\overline{\delta^2(\Delta)}\rangle &\simeq\left\{
    \begin{array}{ll}
          \frac{C_\beta \gamma^2}{2\gamma-1}T^{2\gamma-2}\Delta^2, &~ \gamma>\frac{\alpha}{2}, \\[5pt]
      \frac{C_\beta(\alpha-2\gamma)}{\alpha}T^{\alpha-1}\Delta, & ~\gamma<\frac{\alpha}{2},
    \end{array}\right.
  \end{split}
\end{equation}
where $\beta=\max(2\gamma,\alpha)$ and $C_\beta$ denotes the diffusion coefficient in Eq. \eqref{EAMSDy2}.

\begin{table}
\centering
\caption{Moments of the comoving coordinate $x(t)$ and the physical coordinate $y(t)$ for different forms of the scale factor $a(x,t)$. The notation ``---'' denotes the any form of $a_1(x)$, and $C$ denotes a constant, but different $C$ can take different values.  }\label{table}
\begin{tabular}{|c|c|c|c|}
  \hline
  $a_1(x)$ & --- & --- &  $|x|^\beta$ \\[3pt]
  $a_2(t)$ & $e^{\gamma t}(\gamma>0)$ & $t^{\gamma}(\gamma>\frac{\alpha}{2})$ &  $t^{\gamma}(\gamma<\frac{\alpha}{2})$ \\[3pt]
  $\langle x^2(t)\rangle$ & $C$ & $C$ &  $t^{\frac{\alpha-2\gamma}{1+\beta}}$ \\[3pt]
  $\langle y^2(t)\rangle$ & $e^{2\gamma t}$ & $t^{2\gamma}$ &  $t^\alpha$  \\[3pt]
  $\langle \overline{\delta^2(\Delta)}\rangle$ & $C$ & $\Delta^2$ &  $\Delta$  \\
  \hline
\end{tabular}
\end{table}

\begin{figure*}
  \centering
  \includegraphics[scale=0.55]{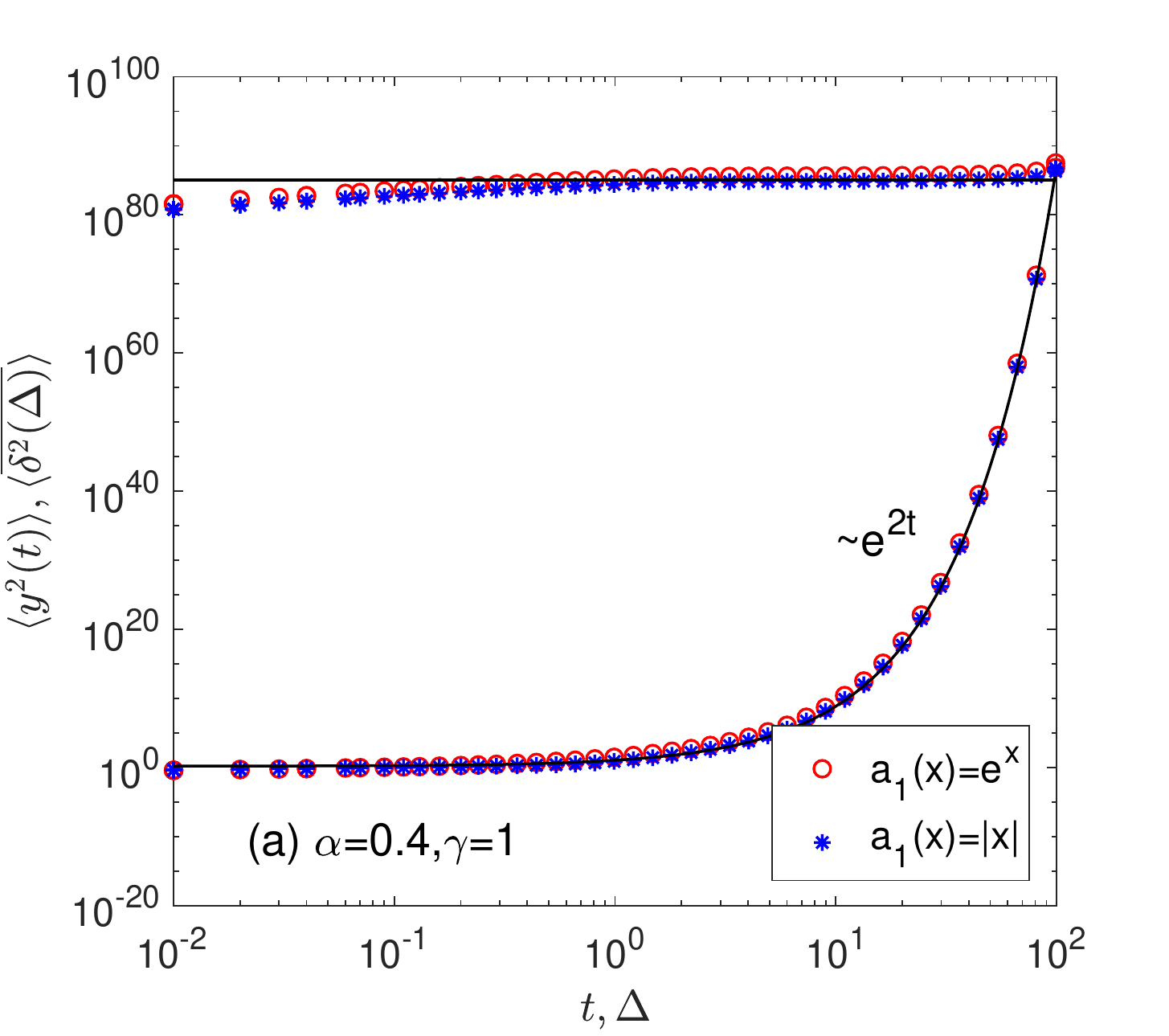}
  \includegraphics[scale=0.55]{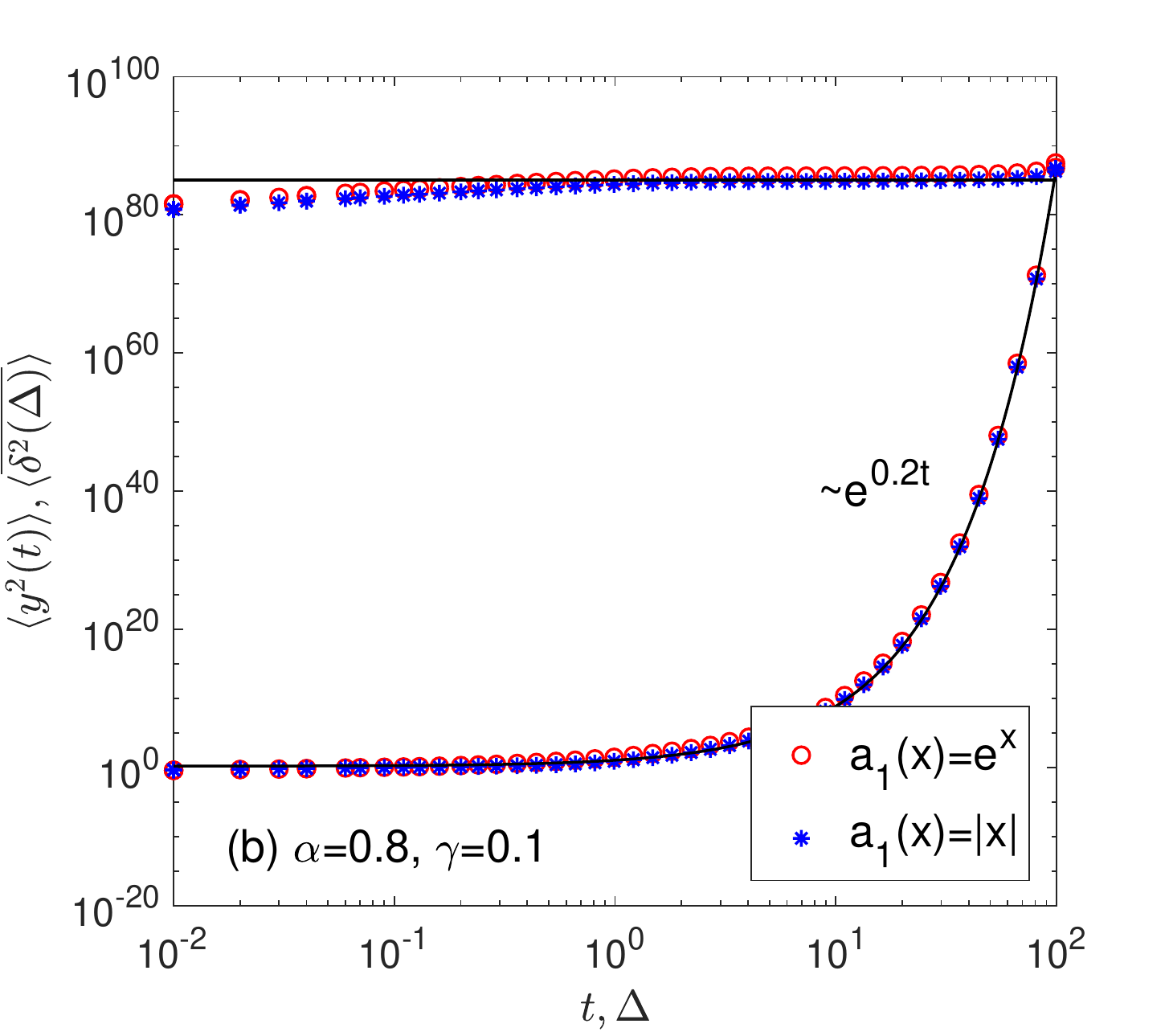}\\
  \includegraphics[scale=0.55]{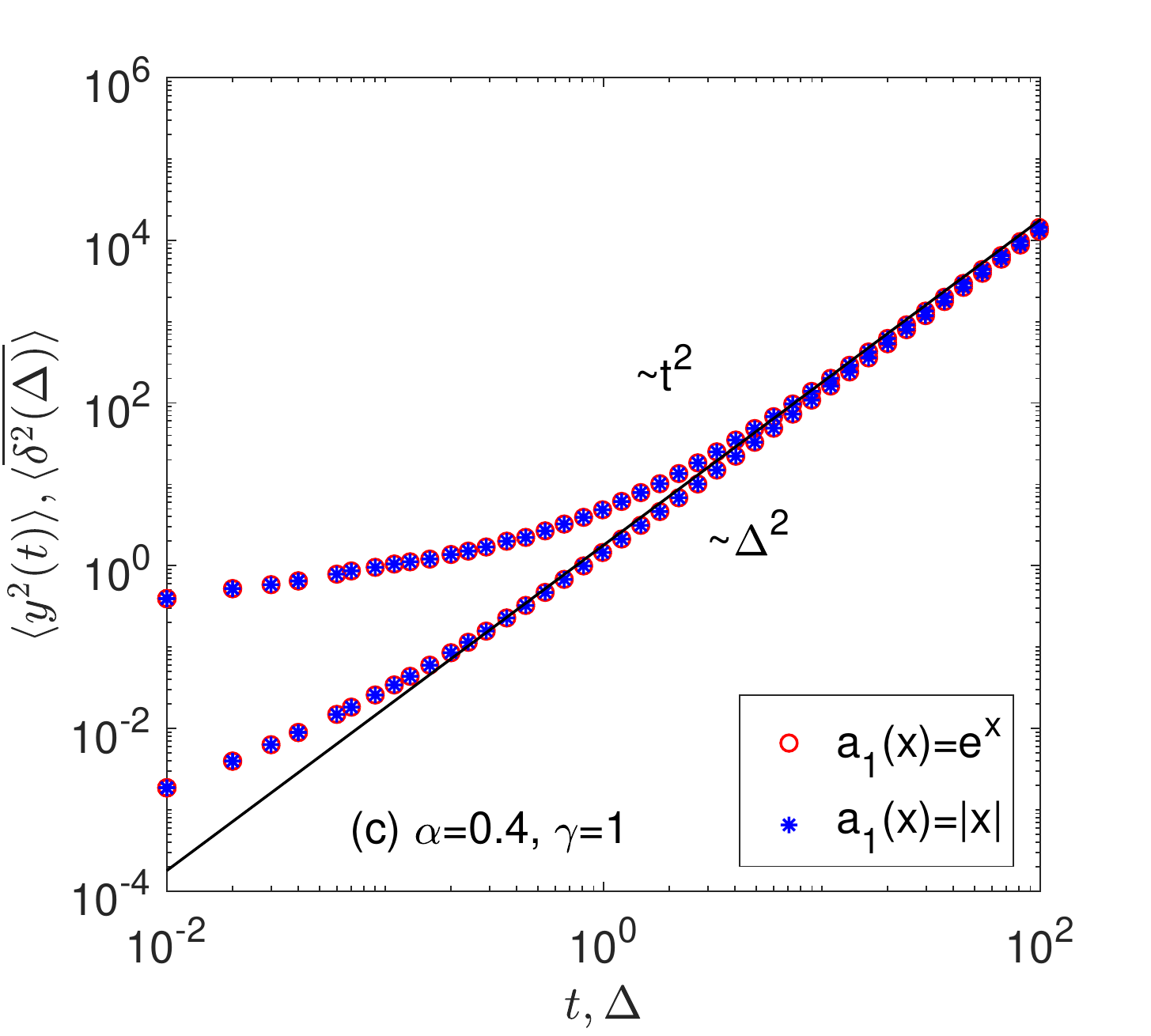}
  \includegraphics[scale=0.55]{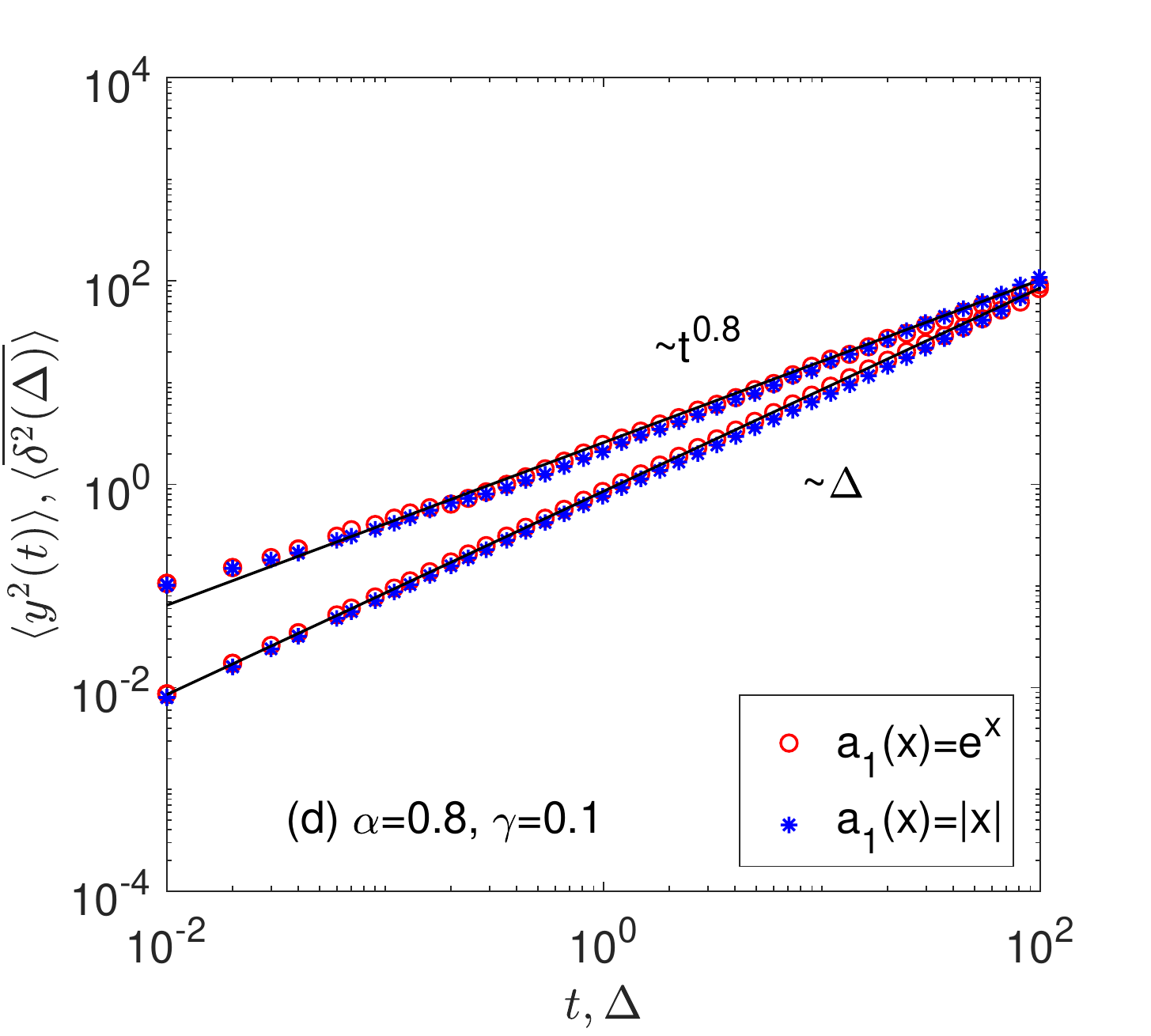}
  \caption{(Color online) EAMSD $\langle y^2(t)\rangle$ and ensemble-averaged TAMSD $\langle\overline{\delta^2(\Delta)}\rangle$ of the physical coordinate $y(t)$ for different scale factor $a(x,t)\simeq a_1(x)a_2(t)$. Two pairs of $\alpha$ and $\gamma$ are taken to correspond to the two cases: $\gamma>\alpha/2$ and $\gamma<\alpha/2$. The upper panels (a) and (b) correspond to the exponential form $a_2(t)=e^{\gamma t}$, while the bottom panels (c) and (d) correspond to the power-law form $a(t)=(1+t/t_0)^\gamma$. The spatial-dependent part $a_1(x)$ is shown in each graph.
  The red circles and blue stars denote the simulated EAMSD and ensemble-averaged TAMSD. The black solid lines represent the correspond theoretical results, which can be found in Eqs. \eqref{EAMSDy1}-\eqref{TAMSDy2} and the Table \ref{table}.
  The simulation markers fit to the theoretical lines well in four panels with different $\alpha$ and $\gamma$.
  Other parameters: $\beta=1$, $D=1$, $T=100$, $t_0=1$, and the number of trajectories used for ensemble is $N=10^3$.}\label{fig2}
\end{figure*}

\begin{figure}
  \centering
  \includegraphics[scale=0.55]{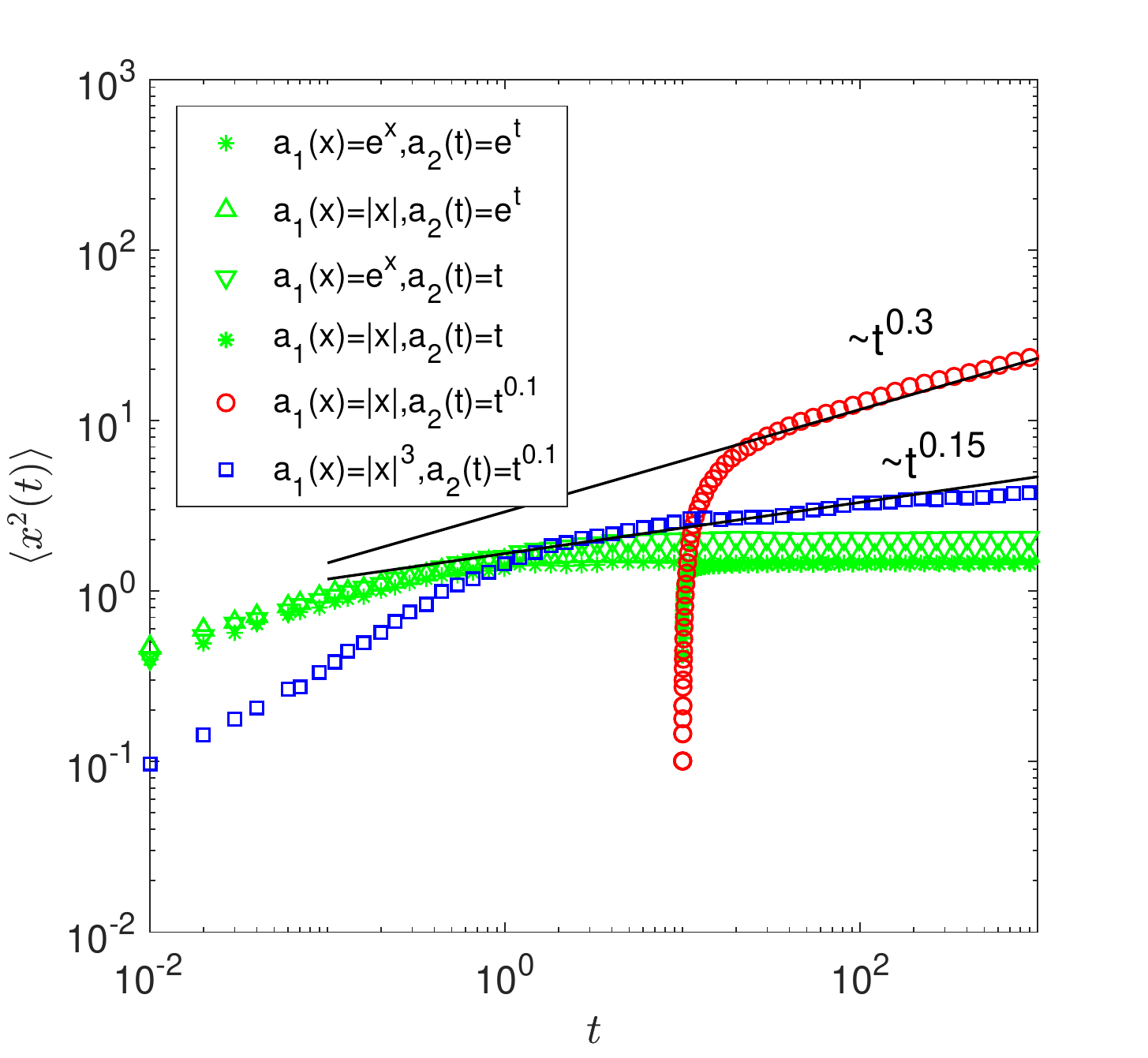}
  \caption{(Color online) EAMSD $\langle x^2(t)\rangle$ of the comoving coordinate $x(t)$ for different scale factor $a(x,t)\simeq a_1(x)a_2(t)$.
  The green markers denotes the simulation results with $\alpha=0.4,\gamma=1$, where the theoretical results tend to a constant as Eq. \eqref{EAMSDx1} and Table \ref{table} shows. The red circles ($\beta=1$) and blue squares ($\beta=3$) represent the simulation results with $\alpha=0.8,\gamma=0.1$ satisfying $\gamma<\alpha/2$. In this case, the EAMSD $\langle x^2(t)\rangle$ depends on the form of $a_1(x)$, where the theoretical results are shown in Eq. \eqref{EAMSDx3} and presented by the solid lines with the increasing rate $t^{0.3}$ and $t^{0.15}$.
  Other parameters: $D=1$, $T=1000$, $t_0=1$, and the number of trajectories used for ensemble is $N=10^3$.}\label{fig3}
\end{figure}

In order to increase the readability, we collect the main results in Secs. \ref{Sec4} and \ref{Sec5}, i.e., the moments with respect to the comoving and physical coordinates, in Table \ref{table}, where different forms of $a_1(x)$ and $a_2(t)$ in the scale factor $a(x,t)$ leads to different moments.  The results for exponential formed $a_2(t)$ with $\gamma>0$ and power-law formed $a_2(t)$ with $\gamma>\alpha/2$ are valid for any form of $a_1(x)$.
In simulations, to guarantee the initial condition of the scale factor, i.e., $a(x,0)=1$, and the asymptotically separable in Eq. \eqref{separable}, we select
\begin{equation}
  a(x,t)=1+a_1(x)(a_2(t)-1),
\end{equation}
where $a_2(t)$ is in the form of Eqs. \eqref{Sim-a2t1} and \eqref{Sim-a2t2}, and $a_1(x)$ is also in exponential form
\begin{equation}
  a_1(x)=e^{\beta x}, \quad \beta>0,
\end{equation}
or power-law form
\begin{equation}
  a_1(x)=|x|^{\beta}, \quad \beta>-1.
\end{equation}
The simulations of the MSDs with respect to the physical coordinate $y(t)$ and the comoving coordinate $x(t)$ are presented in Figs. \ref{fig2} and \ref{fig3}, respectively.

In Fig. \ref{fig2}, we select different forms of the scale factor $a(x,t)\simeq a_1(x)a_2(t)$ and different values of parameters $\alpha$ and $\beta$. Since the theoretical results of the physical coordinate $y(t)$ are independent of the specific form of $a_1(x)$, we collect the exponential form $a_1(x)=e^{\beta x}$ and the power-law form $a_1(x)=|x|^\beta$ in one graph with red circles and blue stars, respectively. It can be found that the two kinds of markers are coincident in each graph.

In Fig. \ref{fig3}, we collect the EAMSD $\langle x^2(t)\rangle$ of the comoving coordinate $x(t)$ in different cases in one graph. As Table \ref{table} shows, the EAMSD tends to a constant when $a_2(t)=e^{\gamma t}(\gamma>0)$ and $a_2(t)=t^\gamma(\gamma>\alpha/2)$, whatever the form of $a_1(x)$ is. So we choose four different scale factor $a(x,t)$, and present the simulation results in green markers, which all tends to a constant. When $a_2(t)=t^\gamma(\gamma<\alpha/2)$, the EAMSD $\langle x^2(t)\rangle$ depends on the specific form of $a_1(x)$. So we take $a_1(x)=|x|^\beta$ with $\beta=1,3$, and show the simulation results in red circles and blue squares, which present the theoretical increasing rate $t^{0.3}$ and $t^{0.15}$, respectively.

\section{Summary}\label{Sec6}

The dynamic mechanism of particle's motion in the expanding medium has been revealed in the framework of CTRW \cite{VotAbadYuste:2017,AngstmannHenryMcGann:2017,VotYuste:2018,Abad-etal:2020,VotAbadMetzlerYuste:2020}, and partly in the framework of Langevin equation for uniformly expanding medium \cite{WangChen:2023}. This paper extends the uniform expansion to the nonuniform expansion where the scale factor $a(x,t)$ is spatial-temporal dependent, and builds the corresponding Langevin picture. For the convenience of theoretical analyses, we assume that the spatial variable $x$ and temporal variable $t$ are separable in the scale factor $a(x,t)$ at long-time limit as Eq. \eqref{separable} shows. In addition to the usual comoving coordinate $x(t)$ and physical coordinate $y(t)$ when solving expanding medium problems, we also introduce the new coordinate $z(t)$, based on which, we can build the relation between the nonuniformly expanding medium and the uniformly expanding one, and further obtain the moments of the comoving and physical coordinates.

Since the scale factor $a(x,t)$ is the key of characterizing the change of the medium, we conduct the analyses by taking specific $a(x,t)$. For the temporal dependent part $a_2(t)$, we mainly consider two forms, the exponential form $a_2(t)=e^{\gamma t}(\gamma>0)$ and the power-law form $a_2(t)=(1+t/t_0)^\gamma(\gamma>0)$.
For exponential-formed $a_2(t)$ and power-law-formed $a_2(t)$ with $\gamma>\alpha/2$, whatever the form of $a_1(x)$ is, the results are similar to the case with uniformly expanding medium in Ref. \cite{WangChen:2023}, i.e., the EAMSDs of the comoving coordinate $\langle x^2(t)\rangle$ are a constant, and the EAMSDs of the physical coordinate $\langle y^2(t)\rangle$ increase as $e^{2\gamma t}$ and $t^{2\gamma}$, respectively, as Table \ref{table} shows.
While for the power-law-formed $a_2(t)$ with $\gamma<\alpha/2$, only the EAMSD and the ensemble-averaged TAMSD of $y(t)$ are the same as the results in Ref. \cite{WangChen:2023}, independent of the specific form of $a_1(x)$. By contrast, the anomalous diffusion exponent of the EAMSD of $x(t)$ depends on the form of $a_1(x)$.
Therefore, the temporal dependent part $a_2(t)$ in the scale factor $a(x,t)$ plays more important role than the spatial dependent part $a_1(x)$.

Due to the essential difficulty of analyzing the general scale factor $a(x,t)$, we assume that $a(x,t)$ is separable with respect to $x$ and $t$ at long-time limit. In detailed theoretical analyses, the temporal dependent part $a_2(t)$ is handled similar to the scaled Brownian motion \cite{LimMuniandy:2002,ThielSokolov:2014,JeonChechkinMetzler:2014,Safdari-etal:2015,SposiniMetzlerOshanin:2019} as Eq. \eqref{modelzt2} shows, while the spatial dependent part $a_1(x)$ in Eq. \eqref{modelxt2} is like the multiplicative noise term in Langevin equation which characterizes a heterogeneous diffusion process \cite{CherstvyChechkinMetzler:2013,CherstvyMetzler:2013,CherstvyMetzler:2014,WangDengChen:2019,LeibovichBarkai:2019}.
More general, the ergodicity breaking and aging of the diffusion processes with spatial-temporal dependent diffusivity are investigated in Ref. \cite{CherstvyMetzler:2015-2}, where the diffusivity is also assumed to be separable. Fortunately, based on our Langevin picture of the particle's motion in the nonuniformly expanding medium, the results for a more general scale factor $a(x,t)$ can be obtained by simulations.

\section*{Acknowledgments}
X.W. acknowledges support by the National Natural Science Foundation of China under Grant No. 12105145 and the Natural Science Foundation of Jiangsu Province under Grant No. BK20210325. Y.C. acknowledges support by the National Natural Science Foundation of China under Grant No. 12205154. W.W. acknowledges supported by the National Natural Science Foundation of China under Grant No. 12105243 and the Zhejiang Province Natural Science Foundation LQ22A050002.

\appendix

\section{Derivation of Eq. \eqref{ZSBM}}\label{App1}

Performing the integral over a small time interval $(t,t+\Delta t)$ over Eq. \eqref{modelzt2}, we get
\begin{equation}\label{App1-1}
  \begin{split}
    z(t+\Delta t)-z(t)&=\sqrt{2D}t^{-\gamma}\Big[B[s(t+\Delta t)]-B[s(t)] \Big] \\
    &\overset{d}{=} \sqrt{2D}t^{-\gamma}\Big[B[s(t+\Delta t)-s(t)] \Big] \\
    &=\sqrt{2D}t^{-\gamma}B[\Delta t \dot{s}(t)]  \\
    &\overset{d}{=} \sqrt{2D}t^{-\gamma}[\dot{s}(t)]^{1/2}B(\Delta t),
  \end{split}
\end{equation}
where we have used the stationary increment property that the distribution of the increment of Brownian motion only depends on the time difference in the second line, the Taylor expansion around the point $t$ in the third line, and the self-similarity of Brownian motion that $B(t)\overset{d}{=}t^{1/2}B(1)$ \cite{BeckerMeerschaertScheffler:2004} in the last line. On the other hand, let $z(t)\overset{d}{=}aB[s(t^\mu)]$. Similar to Eq. \eqref{App1}, we have
\begin{equation}\label{App1-2}
  \begin{split}
   & z(t+\Delta t)-z(t)  \\
    &\overset{d}{=}a\Big[B[s[(t+\Delta t)^\mu]]-B[s(t^\mu)] \Big] \\
    &\overset{d}{=}aB\Big[s[(t+\Delta t)^\mu]-s(t^\mu) \Big] \\
    &\overset{d}{=}aB\Big[t^{\alpha(\mu-1)}\big[s[(1+\Delta t/t)^{\mu-1}(t+\Delta t)]-s(t)\big] \Big] \\
    &=a t^{\alpha(\mu-1)/2}B[\mu\Delta t \dot{s}(t)] \\
    &\overset{d}{=}a t^{\alpha(\mu-1)/2}\mu^{1/2}[\dot{s}(t)]^{1/2}B(\Delta t),
  \end{split}
\end{equation}
where we have used the stationary increment property of Brownian motion in the second line, the self-similarity of the inverse subordinator that $s(t)\overset{d}{=}t^\alpha s(1)$ \cite{BeckerMeerschaertScheffler:2004} in the third line, the Taylor expansion around the point $t$ in the fourth line, and the self-similarity of Brownian motion in the last line.
Comparing Eqs. \eqref{App1-1} and \eqref{App1-2}, we have
\begin{equation}
  \begin{split}
    &\mu=1-2\gamma/\alpha,  \\
    &a=\sqrt{2D}\mu^{-\frac{1}{2}}.
  \end{split}
\end{equation}

\section{Derivation of Eq. \eqref{EAMSDx3}}\label{App2}

If $a_1(x)=|x|^\beta$, then we substitute it into Eq. \eqref{z-x} and obtain $z=|x|^{1+\beta}\mathrm{sgn}(x)/(1+\beta)$, and
\begin{equation}
  x=(1+\beta)^{\frac{1}{1+\beta}}|z|^{\frac{1}{1+\beta}}\mathrm{sgn}(z).
\end{equation}
Therefore, at long-time limit, the EAMSD with respect to the comoving coordinate $x(t)$ is
\begin{equation}
    \langle x^2(t)\rangle \simeq (1+\beta)^{\frac{2}{1+\beta}}\langle |z|^{\frac{2}{1+\beta}}\rangle.
\end{equation}
Based on Eq. \eqref{ztd}, it holds that
\begin{equation}
  \langle x^2(t)\rangle \simeq \tilde{D} t^{\frac{\mu\alpha}{1+\beta}},
\end{equation}
where $\mu=1-2\gamma/\alpha>0$ and
\begin{equation}\label{App2-1}
  \tilde{D}=\left(\frac{2D(1+\beta)^2}{\mu}\right)^{\frac{1}{1+\beta}} \langle|B[s(1)]|^{\frac{2}{1+\beta}}\rangle.
\end{equation}
Now we calculate the fractional moment of random variable $B[s(1)]$ in Eq. \eqref{App2-1}. For convenience, let us consider the subordinated Brownian motion $B[s(t)]$, whose PDF can be written into an integral form as
\begin{equation}\label{App2-2}
  p(x,t)=\int_0^\infty p_0(x,\tau)h(\tau,t)d\tau,
\end{equation}
where $p_0(x,\tau)$ is the PDF of displacement of the standard Brownian motion with Fourier transform $(x\rightarrow k)$ being
\begin{equation}\label{App2-3}
  p_0(k,\tau)=e^{-\tau k^2/2},
\end{equation}
and $h(\tau,t)$ is the PDF of the inverse $\alpha$ stable subordinator with Laplace transform $(t\rightarrow\lambda)$ being
\begin{equation}\label{App2-4}
  h(\tau,\lambda)=\lambda^{\alpha-1}e^{-\tau\lambda^\alpha}.
\end{equation}
Multiplying $|x|^{\frac{2}{1+\beta}}$ on both sides of Eq. \eqref{App2-2}, we obtain
\begin{equation}\label{App2-5}
  \langle |x(t)|^{\frac{2}{1+\beta}}\rangle=\int_0^\infty \langle |x_0(\tau)|^{\frac{2}{1+\beta}}\rangle h(\tau,t)d\tau,
\end{equation}
where
\begin{equation}\label{App2-6}
  \langle |x_0(\tau)|^{\frac{2}{1+\beta}}\rangle = K\tau^{\frac{1}{1+\beta}}
\end{equation}
is the absolute $\frac{2}{1+\beta}$th moment of the standard Brownian motion, and
\begin{equation}\label{App2-7}
  K=2^{\frac{1}{1+\beta}}\pi^{-\frac{1}{2}}\Gamma\left(\frac{1}{1+\beta}+\frac{1}{2}\right).
\end{equation}
Performing the Laplace transform $(t\rightarrow\lambda)$ on Eq. \eqref{App2-5} and substituting Eq. \eqref{App2-4} into it, we obtain
\begin{equation}\label{App2-8}
  \mathcal{L}\{\langle |x(t)|^{\frac{2}{1+\beta}}\rangle\} = K\Gamma\left(\frac{2+\beta}{1+\beta}\right) \lambda^{-1-\frac{\alpha}{1+\beta}}.
\end{equation}
The inverse Laplace transform yields
\begin{equation}\label{App2-9}
  \langle |x(t)|^{\frac{2}{1+\beta}}\rangle = \frac{K\Gamma\left(\frac{1}{1+\beta}\right) t^{\frac{\alpha}{1+\beta}}}
  {\alpha\Gamma\left(\frac{\alpha}{1+\beta}\right)}.
\end{equation}
Taking $t=1$, we obtain the fractional moment of $B[s(1)]$ in Eq. \eqref{App2-1}:
\begin{equation}\label{App2-10}
  \langle |B[s(1)]|^{\frac{2}{1+\beta}}\rangle = \frac{K\Gamma\left(\frac{1}{1+\beta}\right) }
  {\alpha\Gamma\left(\frac{\alpha}{1+\beta}\right)}.
\end{equation}
Substituting Eqs. \eqref{App2-10} and \eqref{App2-7} into Eq. \eqref{App2-1}, the diffusion coefficient equals to
\begin{equation}
  \tilde{D}=2\left(\frac{D(1+\beta)^2}{\mu}\right)^{\frac{1}{1+\beta}}
  \frac{\Gamma\left(\frac{2}{1+\beta}\right)}
  {\Gamma\left(\frac{\alpha}{1+\beta}\right)}.
\end{equation}

\section*{References}
\bibliography{ReferenceW}

\end{document}